\documentclass[12pt]{article}
\pdfoutput=1
\usepackage[utf8]{inputenc}
\usepackage{lmodern}
\usepackage[T1]{fontenc}
\usepackage{amsmath,physics,slashed}
\usepackage{amsfonts}
\usepackage{mathtools}
\usepackage{amssymb,bbm}
\usepackage{dsfont}
\usepackage{subcaption}
\usepackage{braket}
\usepackage{amsthm}
\usepackage{authblk}
\usepackage{xcolor}
\usepackage{cite}
\usepackage[hidelinks]{hyperref}

\usepackage{tikz}
\usetikzlibrary{calc}
\usetikzlibrary{topaths}
\usetikzlibrary{decorations}
\usetikzlibrary{decorations.pathmorphing}
\usetikzlibrary{arrows,decorations.markings,cd}
\usetikzlibrary{calc,arrows,cd,decorations.markings,snakes}
\tikzset{
->-/.style args={#1rotate#2}{decoration={markings, mark=at position #1 with {\arrow[scale=1.5,rotate = #2 ]{stealth}}}, postaction={decorate}}
}
\usetikzlibrary{shapes.geometric}
\usetikzlibrary{knots}
\usepackage{tikz-3dplot}

\hypersetup{
    pdftitle = {Exploring Non-Invertible Symmetries in Free Theories},
    pdfauthor = {Pierluigi Niro, Konstantinos Roumpedakis, Orr Sela}
}

\newcommand{\nc}{\newcommand}
\nc{\rnc}{\renewcommand} 

\rnc{\a}{\alpha}
\rnc{\b}{\beta}
\nc{\g}{\gamma}
\rnc{\d}{\delta}
\nc{\e}{\epsilon}
\nc{\ee}{\varepsilon}
\nc{\z}{\zeta}
\nc{\f}{\phi}
\nc{\m}{\mu}
\nc{\n}{\nu}
\rnc{\r}{\rho}
\rnc{\k}{\kappa}
\rnc{\l}{\lambda}
\nc{\p}{\pi}
\nc{\s}{\sigma}
\rnc{\S}{\Sigma}
\rnc{\t}{\tau}
\nc{\w}{\omega}
\nc{\x}{\chi}
\nc{\F}{\Phi}
\rnc{\L}{\Lambda}
\nc{\cL}{\mathcal L}
\nc{\fb}{{\overline f}}
\nc{\zb}{{\bar z}}
\nc{\A}{\mathcal{A}}
\nc{\LL}{\mathcal{L}}
\nc{\cS}{\mathcal{S}}

\nc{\pd}{\partial}
\nc{\eq}[2]{\begin{align}\label{#1}#2\end{align}}

\nc\ft {{\mathfrak t}}
\nc\fk {{\mathbf k}}
\nc\fg {{\mathfrak g}}
\nc{\T}{\mathcal{T}}

\usepackage[draft,inline,nomargin,index]{fixme}
\FXRegisterAuthor{kr}{akr}{\color{blue}KR}

\begin{document}

\begin{titlepage}
		
		\medskip
		\begin{center} 
			{\Large \bf Exploring Non-Invertible Symmetries in Free Theories}

			\bigskip
			\bigskip
			\bigskip
			
			{\bf Pierluigi Niro, Konstantinos Roumpedakis, Orr Sela\\ }
			\bigskip
			\bigskip
Mani L. Bhaumik Institute for Theoretical Physics, Department of Physics and Astronomy, University of California, Los Angeles, CA 90095, USA\\
			\vskip 5mm				
			
		\end{center}
		
		\bigskip
		\bigskip
	
\begin{abstract}
Symmetries corresponding to local transformations of the fundamental fields that leave the action invariant give rise to (invertible) topological defects, which obey group-like fusion rules. One can construct more general (codimension-one) topological defects by specifying a map between gauge-invariant operators from one side of the defect and such operators on the other side. In this work, we apply such construction to Maxwell theory in four dimensions and to the free compact scalar theory in two dimensions. In the case of Maxwell theory, we show that a topological defect that mixes the field strength $F$ and its Hodge dual $\star F$ can be at most an $SO(2)$ rotation. For rational values of the bulk coupling and the $\theta$-angle we find an explicit defect Lagrangian that realizes values of the $SO(2)$ angle $\varphi$ such that $\cos \varphi$ is also rational. We further determine the action of such defects on Wilson and 't Hooft lines and show that they are in general non-invertible. We repeat the analysis for the free compact scalar $\f$ in two dimensions. In this case we find only four discrete maps: the trivial one, a $\mathbbm{Z}_2$ map $d\f \rightarrow -d\f$, a $\mathcal{T}$-duality-like map $d\f \rightarrow i \star d\f$, and the product of the last two.
\end{abstract}

\end{titlepage}
	
	
\setcounter{tocdepth}{2}	

\tableofcontents

\section{Introduction}
\label{section Introduction}

\subsection{Generalities}

Symmetries are a central tool in the study of Quantum Field Theory. In the last decade the existence of symmetries has become synonymous to the existence of topological operators, i.e.\,operators whose position can be deformed infinitesimally without affecting correlation functions. Usual symmetries are generated by codimension-one topological operators satisfying a group-like fusion rule. This point of view has led to two important generalizations of symmetries. First, the $p$-form symmetries \cite{Gaiotto:2014kfa}, which are generated by topological operators of codimension $p+1$, and by now have become a standard tool for studying QFTs. Second, the so-called {\it non-invertible} symmetries, associated to operators that do not satisfy a group-like fusion rule, and are the subject of this work. 

Non-invertible symmetries are long known to exist in two-dimensional rational CFTs since the work of Verlinde \cite{Verlinde:1988sn} (see also \cite{Petkova:2000ip, Petkova:2001zn,Fuchs:2002cm,Frohlich:2004ef,Frohlich:2006ch, Bhardwaj:2017xup, Chang:2018iay,Ji:2019ugf,Thorngren:2019iar, Thorngren:2021yso,Sharpe:2021srf, Burbano:2021loy}). It was recently realized that non-invertible symmetries also exist in non-conformal theories in two dimensions \cite{Komargodski:2020mxz}, as well as in QFTs in higher dimensions \cite{Nguyen:2021yld, Choi:2021kmx, Kaidi:2021xfk, Choi:2022zal, Cordova:2022ieu, Roumpedakis:2022aik,Bhardwaj:2022yxj,Damia:2022rxw,Arias-Tamargo:2022nlf,Hayashi:2022fkw,Kaidi:2022uux,Antinucci:2022eat, Choi:2022rfe, Choi:2022jqy,Bartsch:2022mpm}, and have turned out to be useful in various contexts \cite{Heidenreich:2021xpr, Cordova:2022rer,Huang:2021zvu, Bashmakov:2022jtl,Benini:2022hzx, Damia:2022bcd, Apruzzi:2022rei,GarciaEtxebarria:2022vzq,Lin:2022dhv,Heckman:2022muc}. However, so far in the literature the existence of non-invertible symmetries has been shown either by using indirect methods or by using methods that are particular to certain classes of theories. One such method requires the presence of a higher-form symmetry which can be gauged along a surface giving rise to a so-called condensation defect \cite{Roumpedakis:2022aik, Lin:2022xod,Bartsch:2022mpm}. Although condensation defects can be constructed in any QFT with a higher-form symmetry, such defects do not act on local operators. Another method for constructing non-invertible defects makes use of dualities \cite{Choi:2021kmx, Choi:2022zal, Choi:2022jqy, Choi:2022rfe}. Although such duality defects can act on local operators, their construction requires the existence of dualities, which is not a generic feature of QFTs.    

In the case of ordinary symmetries, for every transformation that leaves the action invariant there is a topological operator that generates that transformation (assuming there is no anomaly). For continuous $p$-form symmetries, Noether's theorem provides an explicit way to construct that operator as 
\begin{equation}
    U_{g=e^{i\alpha}}[\Sigma_{(d-p+1)}] = \exp \left( i \a \int_{\Sigma_{(d-p-1)}} \star J \right)~,
\end{equation}
where $g$ is the element of the symmetry group, $\Sigma$ is a codimension-$(d-p-1)$ surface, and $J$ is the conserved ($p+1$)-form current ($d\star J=0$). A natural question is whether there is an analog of the above criterion for non-invertible symmetries, which would provide a more systematic way to identify them, as well as to construct the corresponding topological operators explicitly. 

In this work, although we do not attempt to provide a general criterion for the existence of non-invertible symmetries, we put forward a constructive way to analyze them in free theories, such as the four-dimensional Maxwell theory and the two-dimensional theory of a compact scalar. 

In the 4d Maxwell theory, we consider generic defects that act by mixing the field strength $F$ and its Hodge dual $\star F$, and we show that such defects are topological when they act as an $SO(2)$ rotation on these operators. We give a Lagrangian description of these defects for arbitrary rational values of the coupling constants $e^4$ and $\theta/\pi$, realizing the $SO(2)$ rotations with angle $\varphi$ such that $\cos\varphi$ is a rational number. Despite the fact that the action on the local operators $F$ and $\star F$ is invertible, the action on Wilson and 't Hooft lines is generically non-invertible.  

In a similar fashion, we consider topological defects in the 2d free compact scalar theory, which act by mixing $d\phi$ and $\star d\phi$. We show that in this case we only find four types of allowed transformations, corresponding to the trivial one, the $\mathbb{Z}_2$, a $\T$-duality-like transformation for rational values of the squared radius of the target space, and the product of the last two.

Our construction may turn out to be mostly useful for free theories, but we hope that it will also be useful towards developing a criterion for the existence of non-invertible symmetries.

\subsection{Defect Lagrangian}

In this work, we focus on codimension-one defects in free theories, which we construct and study using an explicit Lagrangian description. Let us begin by considering a theory with Lagrangian $\mathcal{L}$, and a defect along a surface $S$ that splits the spacetime into two regions, such that  $M=S^+ \cup S^-$ with $\partial S^- = - \partial S^+ = S$. For simplicity, we choose $S$ to be an infinite flat surface at $x=0$ and denote collectively the fields on the left and right sides of the defect by $\Phi_L$ and $\Phi_R$, respectively. The bulk Lagrangian $\mathcal{L}$ on both sides is the same as we are considering a defect in a single theory and not an interface between two different theories. On the defect itself, the fields from the two sides are related in a non-trivial way determined by the choice of defect Lagrangian $\mathcal{L}_S$. To see this, let us examine the full action in the presence of a defect
\begin{equation}
S= \int_{x<0} \LL (\Phi_L) + \int_{x=0} \LL_S (\Phi_L,\Phi_R,b)+ \int_{x>0} \LL (\Phi_R)  \,,
\label{action generic}
\end{equation}
where $b$ denotes additional dynamical fields living on the defect. Then, the equations of motion for $\Phi_L$, $\Phi_R$, and $b$ on the defect are
\begin{equation}
\begin{split}
     \frac{\pd \LL(\Phi_L)}{\pd d \Phi_L} \bigg\rvert_{S}+\frac{\pd \LL_S(\Phi_L,\Phi_R,b)}{\pd \Phi_L}\bigg\rvert_{S} -(-1)^{p_\Phi}\,d\,\frac{\pd \LL_S(\Phi_L,\Phi_R,b))}{\pd d \Phi_L}\bigg\rvert_{S} & = 0\,,\\
     \frac{\pd \LL(\Phi_R)}{\pd d \Phi_R}\bigg\rvert_{S} -\frac{\pd \LL_S(\Phi_L,\Phi_R,b)}{\pd \Phi_R}\bigg\rvert_{S} +(-1)^{p_\Phi}\,d\,\frac{\pd \LL_S(\Phi_L,\Phi_R,b))}{\pd d \Phi_R}\bigg\rvert_{S} & = 0\,,\\
     \frac{\pd \LL_S(\Phi_L,\Phi_R,b)}{\pd b}\bigg\rvert_{S} -(-1)^{p_b}\,d\,\frac{\pd \LL_S(\Phi_L,\f_R,b))}{\pd db}\bigg\rvert_{S} & = 0\,,
\end{split}
\label{eom defect general}
\end{equation}
where $p_\Phi$ and $p_b$ are the form degrees of $\Phi_{L,R}$ and $b$, respectively. Note that the first term in the first two equations results from integrating by parts the variation of the bulk Lagrangian, while all the other terms correspond to the variation of the defect Lagrangian. We see that for a given theory with Lagrangian $\mathcal{L}$, the above relations \eqref{eom defect general} between $\Phi_L$ and $\Phi_R$ are determined by the defect Lagrangian $\mathcal{L}_S$. These relations do not necessarily correspond to a transformation which is a symmetry of the action, acting on the fundamental fields locally. In the latter case, that leads to an invertible defect. However, in general the map between $\Phi_L$ and $\Phi_R$  defines a non-invertible defect. 
%

Not every defect Lagrangian $\mathcal{L}_S$ leads to a topological defect. For this, the energy-momentum tensor has to be conserved in the presence of the defect. To see what this implies for $\mathcal{L}_S$, let us first rewrite the action as
\begin{equation}
S= \int_{M} \big(\LL (\Phi_L) \Theta(-x) + \LL_S (\Phi_L,\Phi_R,b) \delta(x)+ \LL (\Phi_R)\Theta(x)\big) \,,
\label{generic action with thetas}
\end{equation}
where $\Theta(x)$ and $\delta(x)$ are the usual theta and delta functions, respectively. We can now compute the total energy-momentum tensor from the variation of \eqref{generic action with thetas} with respect to the metric. Assuming that the defect Lagrangian does not depend on the metric, in order for the energy-momentum tensor to be conserved, it should obey
\begin{equation}
n^\mu (T_{\mu\nu}(\Phi_L)-T_{\mu\nu}(\Phi_R))\big|_{S}=0 \,.
\label{matchingtensor}
\end{equation}
This condition is satisfied only for specific relations between $\Phi_L$ and $\Phi_R$. Hence, the relations \eqref{eom defect general} serve as a constraint to be imposed on $\mathcal{L}_S$ to obtain a topological defect.\footnote{Under the assumption that the defect Lagrangian does not depend on the metric, the condition \eqref{matchingtensor} is equivalent to the vanishing of the displacement operator, defined by $D_\m(\s^a)=\frac{\delta S}{\delta X^\m(\s^a)}$, where $X^\m(\s^a)$ is the defect embedding and $\s^a$ the coordinates parametrizing the defect. See e.g. \cite{Cuomo:2021cnb} for a recent discussion.}

In this note, instead of starting with a given defect Lagrangian,
our strategy will be first to determine the relations between $\Phi_L$ and $\Phi_R$ on the defect such that equation \eqref{matchingtensor} is satisfied. Once we find such relations, we have a candidate for a topological defect (which can be either invertible or not) that generates such action on local operators. Then, our goal will be to give an explicit construction of such topological defect by finding a defect Lagrangian that leads to those relations through \eqref{eom defect general}. 

Since our discussion so far was entirely classical, one might wonder whether the defects constructed in this way are also realized in the quantum theory. If they are not, that would be the analog of an ABJ anomaly for generic topological defects. For a general Lagrangian we cannot provide a sufficient condition that guarantees that they survive quantum corrections. Below, we analyze usual (group-like) symmetries using this approach and show that in the presence of an anomaly the defect is still topological, but it becomes an interface between two different theories. Nevertheless, in the example we are considering, namely Maxwell theory, defects of the type we discuss can also be constructed by gauging discrete subgroups of the one-form symmetry combined with $SL(2,\mathbbm{Z})$ duality transformations on half-space (see appendix \ref{app: Gauging}). This can be done at the quantum level and shows that such defects exist in the quantum theory as well.

\subsection{Ordinary Symmetries}

Let us examine how ordinary symmetries (continuous or not) are described in this Lagrangian approach. Consider a symmetry of the action, transforming the fundamental fields as
\begin{equation}
    \Phi \rightarrow U(\Phi)\,.
\end{equation}
We then schematically define the corresponding defect as 
\begin{equation}
    S = \int_{S^-} L(\Phi_L) + \int_{S} b \wedge (\Phi_L - U(\Phi_R)) +\int_{S^+} L(\Phi_R) \,,
    \label{cont sym}
\end{equation}
where $b$ is a Lagrange-multiplier field that enforces the relation $\Phi_L |_S = U(\Phi_R)|_S$. To see how the defect acts on operators, let us insert it inside a correlation function. Then, operator insertions on the left side of the defect will depend on $\Phi_L$, while operators on the right will depend on $\Phi_R$. We can now trivialize the defect Lagrangian by redefining the bulk fields only on one side of the defect. For example, we can redefine 
\begin{equation}
    \Phi_R \rightarrow U^{-1}(\Phi_R) \label{right transf} \,,
\end{equation}
which will transform all the operator insertions constructed from $\Phi_R$ on the right side of the defect. If there is no anomaly (see below) the defect action now becomes 
\begin{equation}
    \int_{S} b \wedge (\Phi_L -\Phi_R)\,, \label{trivial defect}
\end{equation}
which is simply the trivial defect identifying $\Phi_L$ and $\Phi_R$ on it. Hence, the insertion of \eqref{cont sym} is equivalent to transforming the operators on one side of the defect, leading to the usual Ward identities.

Notice that to trivialize the defect, we assumed that \eqref{right transf} can be done at the quantum level, which requires the absence of anomalies. Instead, if there is an anomaly then the redefinition \eqref{right transf} will produce a bulk term of the form
\begin{equation}
    \int_{S^+} \mathcal{A}(\Phi_R)\,,
    \label{anomaly}
\end{equation}
where $\mathcal{A}(\Phi_R)$ is the anomaly coming from the transformation of the functional measure in the path integral. In this case, the defect becomes an interface between two different theories, one with and one without the term \eqref{anomaly} in the action.

\subsection{Organization}
This work is organized as follows.
In section \ref{section Maxwell defects}, we consider 4d Maxwell theory and study transformations acting linearly on $F$ and $\star F$. We show that such transformations can be at most an $SO(2)$ rotation in order for the corresponding defect to be topological. For rational values of the bulk coupling constant $e^4$ and $\theta/\pi$, we show that there is a defect Lagrangian which realizes rational values of $\cos\varphi$, where $\varphi$ is the $SO(2)$ rotation angle. 

In section \ref{section action on lines}, we study the action of the above defects on line operators. We show that, in general, a loop operator dragged across the defect becomes a disc operator, which is determined by the relation between left and right fields imposed on the defect. We then turn to the more interesting case of line operators piercing the defect. In this case, the operators at the junction are monopole operators and the (generically non-invertible) action on lines is determined by their quantum numbers. 

In section \ref{section scalar defects}, we repeat the discussion of section \ref{section Maxwell defects} in the case of the 2d compact scalar theory, by considering transformations that mix $d\f$ and $\star d\f$. This case is not as rich as the Maxwell one and we only find four discrete types of transformations: the trivial one, a $\mathbb{Z}_2$, a $\T$-duality-like transformation, and the product of the last two. 

Several appendices contain additional material, which makes contact with previous results in the literature and supplements the main text. In appendix \ref{app: Gauging}, we show how the defects we construct for Maxwell theory in section \ref{section Maxwell defects} can also be constructed using gauging and dualities. In appendix \ref{app: CS defect}, we review the action of condensation defects on line operators in 3d Chern-Simons theory using the approach discussed in this note. In appendix \ref{app: S dual frame}, we show how the action on lines in Maxwell theory of a particular defect we investigated in section \ref{section action on lines} fits with the dualization procedure discussed in \cite{Witten:1995gf}, and how it can be analyzed using it. In appendix \ref{app: generic defects}, we explain in more detail how a generic defect of the type considered in section \ref{section Maxwell defects} acts on lines. Finally, in appendix \ref{app: T fusion} we calculate the fusion of two $\T$-duality-like defects using their Lagrangian description.

\section{Topological defects in 4d Maxwell theory}
\label{section Maxwell defects}
Consider the Maxwell theory with a theta term in four dimensions, with a Euclidean Lagrangian 
\begin{equation}
\LL (e^2,\theta;A)= \frac{1}{4\pi e^2} F \wedge \star F + \frac{i\theta}{8\pi^2} F \wedge F \,,
\end{equation}
where $F=dA$ is the field strength. The symmetry of the theory includes a $\mathbb{Z}_2$ charge conjugation symmetry and two continuous one-form symmetries $U(1)_e$ and $U(1)_m$ which measure the electric and magnetic charges
\begin{equation}
n = \int_{\Sigma_2} \frac{\tilde{F}}{2\pi}\,,
\qquad m = \int_{\Sigma_2} \frac{F}{2\pi}~,
\label{emcharges}
\end{equation}
of Wilson and 't Hooft lines
\begin{equation}
W_n(\g) = e^{i n \oint_\g A} \,, \qquad H_m(\g) = e^{-i m \oint_\g \tilde{A}} \,,
\end{equation}
where we have defined the dual field strength 
\begin{equation}
    \tilde{F} = d \tilde{A}= -\frac{i}{e^2} \star F + \frac{\theta}{2\pi} F~.
    \label{dualF}
 \end{equation}
In addition, Maxwell theory enjoys $\cS$-duality which exchanges $F$ and $\tilde{F}$ while acting on the coupling as $\tau\rightarrow -1/\tau$, where $\tau=\frac{\theta}{2\pi}+\frac{i}{e^2}$. At the self-dual point $\t=i$, it becomes a global symmetry exchanging $F$ and $\star F$. In the following, we study more general (non-invertible) topological defects implementing this transformation away from the self-dual point, which we refer to as $\cS$ defects. Such defects have already been constructed in \cite{Choi:2021kmx} for integer values of $i\t$, and we will discuss below how to realize them also for generic rational values of $e^4$ and $\theta/\pi$. In fact, our goal will be to study topological defects in Maxwell theory that implement even more general transformations between $F$ and $\star{F}$. Let us emphasize that in this discussion we consider defects in the same theory and not interfaces between different theories. Hence, the coupling $\t$ is the same on both sides of the defect.

As explained in section \ref{section Introduction}, we define a codimension-one defect by the action\footnote{Due to the $2\pi$ periodicity of $\theta$, one could in principle allow for a $2\pi \mathbb{Z}$ difference between the left and the right theta angles. However, such a difference can be reabsorbed in integer shifts of the levels of the 3d Chern-Simons terms for the $A_R$ and $A_L$ fields on the defect. Since such terms will be present in our defect Lagrangian description, we assume that $\theta_L=\theta_R$ without loss of generality.}
\begin{equation}
S=\int_{S^-} \LL (e^2,\theta;A_L) + \int_{S^+} \LL (e^2,\theta;A_R) + \int_{S} \LL_S (A_L,A_R,b) \,.
\label{actionmaxwell}
\end{equation}
Before we specify the defect Lagrangian, let us ask what is the most general transformation that acts linearly on $F$ and $\star F$. To this end, we consider a transformation of the form
\begin{align}
    F_L |_S &= \alpha \, F_R|_S + \beta \; i \star F_R |_S\,,\label{transformation1}\\
    i\star F_L|_S &= \gamma \, F_R|_S + \delta \; i \star F_R|_S \,,
\label{transformation2}
\end{align}
where $\alpha$, $\beta$, $\gamma$ and $\delta$ are real parameters. Assuming that the defect Lagrangian does not depend on the metric, for the defect to be topological the energy-momentum tensor
\begin{equation}
T_{\mu\nu}(A) = \frac{1}{2\pi e^2} \left( -F_{\mu\alpha}F^\alpha_{\,\nu} - \frac{1}{4} g_{\mu\nu} F_{\alpha\beta}F^{\alpha\beta} \right),
\label{tensormaxwell}
\end{equation}
has to satisfy \eqref{matchingtensor}. Using this explicit expression for $T_{\mu\nu}$, we get that the transformation generated by a topological defect is at most an $SO(2)$ rotation,
\begin{equation}
    \begin{pmatrix}
F_L|_S \\
i\star F_L |_S 
\end{pmatrix}
=
    \begin{pmatrix}
\cos\varphi & \sin\varphi \\
-\sin\varphi  & \cos\varphi  
\end{pmatrix}
    \begin{pmatrix}
F_R|_S  \\
i\star F_R|_S 
\end{pmatrix}\,.
\label{Maxwell sewing}
\end{equation}
The special cases $\varphi = \pi$ and $\varphi =\pi/2$ correspond to the $\mathbb{Z}_2$ and $\cS$ defects, respectively. Another subset of the above defects with a rational $\cos \varphi = - \frac{\theta}{2\pi} $ and $\frac{1}{e^2}=\sin \varphi$ has previously been constructed in \cite{Choi:2022rfe} by gauging a subgroup of the one-form symmetry accompanied with a duality transformation in half-space. In addition, the defect corresponding to $\tau=e^{2\pi i/3}N$ (where $N\in\mathbb{N}$) and $\varphi=-2\pi/3$, which is an example of a triality defect, has been constructed in \cite{Choi:2022zal} using a similar method. 
We next turn to discuss explicit realizations of the defect Lagrangian $\LL_S$ for rational values of $\cos\varphi$, $e^4$ and $\theta/\pi$ (we discuss the construction of such defects using gauging and dualities in appendix \ref{app: Gauging}). It would be interesting if it is possible to realize the whole $SO(2)$ transformations with also irrational values of $\cos\varphi$. We leave that for future work.

\subsection{\texorpdfstring{$\sin\varphi=0$}{sin φ = 0}}
\label{trivial and Z2 section}
Let us first consider two special values for $\varphi$, namely $\varphi=0$ corresponding to the trivial $SO(2)$ element ($F_L|_S =F_R|_S $, $\star F_L|_S = \star F_R|_S $), and $\varphi=\pi$ corresponding to the non-trivial $\mathbb{Z}_2$ element ($F_L|_S =-F_R|_S $, $\star F_L|_S =-\star F_R|_S $). 

There are multiple non-trivial defects corresponding to the trivial $SO(2)$ element. In the literature these are known as condensation defects \cite{Roumpedakis:2022aik, Choi:2022zal}. Although they act trivially on local operators, they act non-trivially on non-local operators like Wilson and 't Hooft lines. In these cases the gauge fields on the two sides of the bulk differ only by a discrete gauge field (with zero field strength) and \eqref{matchingtensor} is trivially satisfied. They arise from gauging a non-anomalous\footnote{In order for a discrete subgroup of a one-form symmetry to be gaugable along a codimension-one surface, it needs to be free from 1-anomalies \cite{Roumpedakis:2022aik}.} discrete subgroup of the electric or magnetic one-form symmetry  \cite{Choi:2021kmx,Choi:2022jqy,Choi:2022zal,Choi:2022rfe}.

An example of a condensation defect \cite{Roumpedakis:2022aik,Choi:2022zal} for arbitrary values of $e^2$ and $\theta$ is realized by the defect Lagrangian 
\begin{equation}
\LL_S = \frac{i \kappa}{2\pi} a \wedge (dA_R - dA_L)~,
\label{condensation Lagrangian}
\end{equation}
where we have introduced an auxiliary $U(1)$ gauge field $a$ living only on the defect, and $\kappa\in\mathbb{Z}$ for gauge invariance (without loss of generality we can choose $\kappa$ to be positive). The equation of motion for $a$ enforces the condition $F_L|_S=F_R|_S$, while the equations of motion for the bulk gauge fields read
\begin{equation}
\frac{\kappa}{2\pi} da = -\frac{i}{2\pi e^2} \star F_R|_S + \frac{\theta}{4\pi^2} F_R|_S= -\frac{i}{2\pi e^2} \star F_L|_S + \frac{\theta}{4\pi^2} F_L|_S\,.
\end{equation}
Combining them, we also get $\star F_L|_S=\star F_R|_S$. Integrating the above equation over a closed two-surface on the defect we get
\begin{equation}
\kappa \, m = n_R = n_L \,,
\label{charges of condensation monopole}
\end{equation}
where the integer $m$ measures the flux of $a$ and $n_{R,L}$ are the one-form electric charges defined in \eqref{emcharges}. Note that for $\kappa \neq 1$ this imposes the constraint on the charges $n_{R,L}$ to be multiples of $\kappa$. As we will see in section \ref{section action on lines},  this implies that although the defect acts trivially on local operators it actually induces a non-trivial transformation on line operators. On the other hand, for $\kappa=1$ equation \eqref{charges of condensation monopole} does not impose any constraint and indeed in this case the defect is trivial. 

Similarly, the non-trivial $\mathbbm{Z}_2$ element of the $SO(2)$ can be realized by the defect Lagrangian  
\begin{equation}
\LL_S = \frac{i \kappa}{2\pi} a \wedge (dA_R + dA_L) \,.
\label{non simple C}
\end{equation}
For $\kappa=1$, it induces the usual (invertible) charge conjugation symmetry.  For $\kappa \neq 1$ it is a product of the charge conjugation and a condensation defect. To see this, consider two defects on two surfaces $S$ and $S+\d S$, one generating the charge conjugation and the other a condensation defect as in \eqref{condensation Lagrangian}. The combined defect action is given by 
\begin{equation}
     \frac{i}{2\pi}\int_S a_1 \wedge (dA_I + dA_L) +\frac{i\kappa}{2\pi}\int_{S+\d S} a_2 \wedge (dA_R - dA_I)\,,
\end{equation}
where $A_I$ is the gauge field in the strip between the two surfaces. Taking the limit $\d S \rightarrow 0$ we get 
\begin{equation}
     \int_S \left( \frac{i}{2\pi} a_1 \wedge (dA_I + dA_L) +\frac{i\kappa}{2\pi} a_2 \wedge (dA_R - dA_I) \right),
\end{equation}
and after redefining $a_1 \rightarrow a_1 + \kappa a_2$ and $A_I \rightarrow A_I -A_L$ we find 
\begin{equation}
     \int_S \left( \frac{i}{2\pi} a_1 \wedge dA_I +\frac{i\kappa}{2\pi} a_2 \wedge (dA_R + dA_L) \right).
\end{equation}
The first term corresponds to a trivial decoupled TQFT and can be dropped. Hence, we see that the product of the usual charge conjugation defect and the condensation defect \eqref{condensation Lagrangian}  is equal to \eqref{non simple C}.

\subsection{\texorpdfstring{$\sin\varphi \neq 0$}{sin φ not 0}  }
\label{general case section}
Let us now construct defects that generate transformations with $\sin\varphi\neq 0$. Those transformations arise as boundary conditions which have to be imposed on the theory in order to satisfy a well-defined variational principle. Hence, we need to impose that if we vary $A_{L,R}\rightarrow A_{L,R} + \delta A_{L,R}$ the corresponding variation of the defect action cancels the contribution coming from the bulk, when \eqref{Maxwell sewing} are satisfied. The bulk variation reads
\begin{equation}
\delta S_{bulk} = \int_S \left( \frac{1}{2\pi e^2}  \d A_L \wedge \star  dA_L + \frac{i\theta}{4 \pi^2}\d A_L  \wedge dA_L - \frac{1}{2\pi e^2}  \d A_R \wedge \star  dA_R - \frac{i\theta}{4 \pi^2}\d A_R  \wedge dA_R\right)\,, 
\end{equation}
where we have ignored the contributions on $S^+$ and $S^-$, which lead to the bulk equations of motion $d\star F_L = d\star F_R=0$. Using \eqref{Maxwell sewing} to express $\star F_L$ and $\star F_R$ in terms of $F_L$ and $F_R$ we get
\begin{equation}
\delta S_{bulk} = \int_S \left(-\frac{i N_L}{2\pi}  \, \delta A_L \wedge dA_L - \frac{i N_R}{2\pi} \delta A_R \wedge dA_R +\frac{i N}{2\pi} (\delta A_L \wedge dA_R+\delta A_R\wedge dA_L) \right),
\label{boundary variation}
\end{equation}
where
\begin{equation}
N_L=\frac{1}{e^2\tan\varphi}-\frac{\theta}{2\pi} \,, \qquad N_R=\frac{1}{e^2\tan\varphi}+\frac{\theta}{2\pi} \,, \qquad N=\frac{1}{e^2\sin\varphi} \,.  \label{N in terms on couplings}
\end{equation}
Naively, a defect Lagrangian whose variation cancels \eqref{boundary variation} is
\begin{equation}
\LL_S = \frac{i N_L}{4\pi} A_L \wedge dA_L + \frac{i N_R}{4\pi} A_R \wedge dA_R -\frac{i N}{2\pi} A_L \wedge dA_R \,.
\label{naivelagrangian}
\end{equation}
However, gauge invariance requires $N_L$, $N_R$, and $N$ to be integers.
Together with the relations \eqref{N in terms on couplings}, this implies that only some special values of the angle $\varphi$ can be realized for given values of the bulk couplings $e^2$ and $\theta$.

In the rest of this section, we attempt to realize any rational value of $N_L$, $N_R$, and $N$. First, let us split $N_L$, $N_R$, and $N$ as
\begin{equation}
N_L = k_L + \frac{p_L}{q_L} \,,\qquad N_R = k_R + \frac{p_R}{q_R} \,, \qquad N = k + \frac{p_N}{q_N} \,,
\end{equation}
where $k_L,k_R$ and $k$ are integers chosen such that $\frac{p_i}{q_i}\in [0,1)$, and gcd$(p_i,q_i)=1$ for every $i=L,R,N$. The integer part of the variation \eqref{boundary variation} can be canceled by \eqref{naivelagrangian} with the $N_i$'s replaced by their integer parts $k_i$. In order to realize the non-integer part of the defect Lagrangian we need to introduce additional dynamical fields living on the defect. One can easily write down a Lagrangian with the right variation, but a generic choice a priori might lead to a non-simple defect, such as a direct sum of more elementary defects. 

To avoid this, we make use of the minimal TQFT $\A^{q,p}$ in three dimensions with one-form symmetry $\mathbb{Z}_q$ and anomaly labeled by $p$ ($p \sim p+ q$) \cite{Hsin:2018vcg}. Because of the anomaly, when the $\A^{q,p}$ theory is coupled to a background two-form gauge field $B$ for the one-form symmetry, it is not invariant under background gauge transformations $B\rightarrow B+ d\l$. For an infinitesimal transformation, it changes as 
\begin{equation}
    \int_M \A^{q,p}[B + d\l] = \int_M \A^{q,p}[B] - \frac{iqp}{2\pi} \int_M  \l \wedge B \,.
\end{equation}
In our case, we can couple the bulk Maxwell theory to an $\A^{q,p}[B]$ TQFT on the defect by activating its background $\mathbb{Z}_q$ gauge field $B$ using the bulk field strengths $dA_L$ and $dA_R$. More specifically, we can choose 
\begin{equation}
    B = \frac{l \; dA_L + r \; dA_R}{q} \,,
\end{equation}
for some integers $l$ and $r$. 

One can show that coupling to a single minimal TQFT is not enough to realize all the rational values of $e^4$ and $\theta/\pi$. Instead, we couple the theory to two  minimal TQFTs and consider the defect Lagrangian 
\begin{equation}
\LL_S = \frac{i k_L}{4\pi} A_L \wedge dA_L + \frac{i k_R}{4\pi} A_R \wedge dA_R -\frac{i k}{2\pi} A_L \wedge dA_R + \LL_\A \,,
\label{generic Ls}
\end{equation}
with
\begin{equation}
\LL_\A = \A^{q_1, p_1}\left[\frac{l_1 \, dA_L + r_1 \, dA_R}{q_1}\right] + \A^{q_2, p_2}\left[\frac{l_2 \, dA_L + r_2 \, dA_R}{q_2}\right] \,, 
\label{Defect L}
\end{equation}
where $l_i,r_i$ are integers and $p_i$ is defined modulo $q_i$, with gcd$(p_i,q_i)=1$ for $i=1,2$. The parameters in the above action are related to the $N_i$'s by 
\begin{equation}
\begin{split}
    N_L &= k_L -\frac{p_1}{q_1} l_1^2-\frac{p_2}{q_2} l_2^2  \,, \\
    N_R &= k_R  -\frac{p_1}{q_1} r_1^2-\frac{p_2}{q_2} r_2^2 \,, \\
    N &= k  +\frac{p_1}{q_1} l_1 r_1+\frac{p_2}{q_2} l_2r_2 \,.
\end{split}
\label{Ns in terms of plrn}
\end{equation}
In this way, any rational value for the $N_i$'s is allowed. We can further invert equations \eqref{N in terms on couplings} to get
\begin{align}
\frac{\theta}{2\pi} &= \frac{N_R-N_L}{2} \,, \label{constraint1}\\
\frac{1}{e^4} &= N^2 - \left(\frac{N_R + N_L}{2}\right)^2 \label{constraint2}\,.
\end{align}
For a theory with given rational coupling constants $e^4$ and $\theta/\pi$, the above can be viewed as constraints on the allowed values of parameters in \eqref{Defect L}. Given the above constraints, one can construct defects that realize elements of the $SO(2)$ in \eqref{transformation2}, labeled by an angle $\varphi$ that satisfies
\begin{equation}
\cos\varphi= \frac{N_R+N_L}{2N} \,.
\end{equation}
It is easy to see that there are several choices of the parameters in \eqref{Defect L} (and hence several defects) that lead to the same $SO(2)$ transformation (for fixed bulk couplings). However, this does not mean that such defects act the same on non-local operators. For instance, one can find examples of two defects $D_\varphi$ and $D'_\varphi$ that give rise to the same angle $\varphi$ and are related by a condensation defect $\mathcal{C}$. This means that $D_\varphi =  (Z) \; \mathcal{C} \times D'_\varphi$, where $Z$ is the partition function of a decoupled TQFT. This suggests that different choices of the parameters  leading to the same $SO(2)$ angle are related by fusing with a condensation defect. However, we were not able to show this for a general angle nor that fusing with a generic condensation defect produces another defect of the form \eqref{Defect L}. 

As an example, let us consider the case of the $\cS$ defect $\varphi=\pi/2$. Given arbitrary rational values of $e^2$ and $\theta/\pi$, such a defect is realized by taking $N_R=-N_L=\frac{\theta}{2\pi}$ and $N=1/e^2$.
Another special example is the case where $\theta=0$. In this case we have $N_R=N_L \equiv M$ and
\begin{equation}
\cos\varphi=\frac{M}{N} \,, \qquad \frac{1}{e^4}=N^2-M^2 \,.
\end{equation}
Given an arbitrary rational value for $e^4$ (since any rational can be written as the difference of two squared rationals $N$ and $M$), all $\varphi$ rotations satisfying the equations above can be realized. In particular, the $\cS$ defect discussed above can be realized at $\theta=0$ by choosing $M=0$ and $N=1/e^2$.

\section{Action on Wilson and 't Hooft lines}
\label{section action on lines}
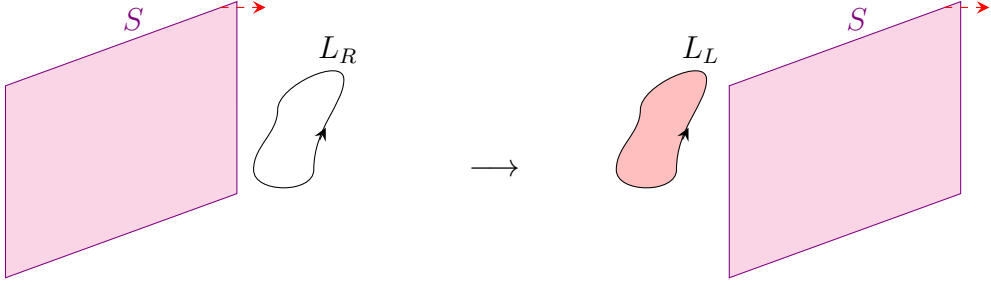
\begin{figure}
	\centering
	\begin{tikzpicture}[scale = 0.8]
		\node[	trapezium, 
		draw = violet,
		minimum width=3.2cm,
		trapezium left angle=70, 
		trapezium right angle=110, 
		rotate = 20, 
		trapezium stretches=false,
		minimum height=2.4cm, 
		fill = magenta!20,
		]
		at (0.8,0.5) {};
		
		\draw[] (3,0) .. controls +(0,-0.4) and +(0,-0.4) .. (4,0).. controls +(0,0.7) and +(0,-0.4) .. (4.5,1.5).. controls +(0,0.4) and +(0,0.4) .. (3.4,1).. controls +(0,-0.4) and +(0,0.4) .. (3,0);
		\draw[->-=1 rotate 0, dashed] (4.1,0.5) -- (4.2,0.7);
		\node[] at (4.4,2) {$L_R$};
		\node[color = violet] at (1,2.5) {$S$};
		\draw[->-=1 rotate 0, color = red, dashed] (2.45,2.7) -- (3.2,2.7);
		\node[] at (7,0) {$\longrightarrow$};
	\end{tikzpicture}
	\hskip 1 cm
	\begin{tikzpicture}[scale = 0.8]
		\node[	trapezium, 
		draw = violet, 
		minimum width=3.2cm,
		trapezium left angle=70, 
		trapezium right angle=110, 
		rotate = 20, 
		trapezium stretches=false,
		minimum height=2.4cm, 
		fill = magenta!20,
		]
		at (6.8,0.5) {};
		\draw[fill = pink] (3,0) .. controls +(0,-0.4) and +(0,-0.4) .. (4,0).. controls +(0,0.7) and +(0,-0.4) .. (4.5,1.5).. controls +(0,0.4) and +(0,0.4) .. (3.4,1).. controls +(0,-0.4) and +(0,0.4) .. (3,0);
		\draw[->-=1 rotate 0, dashed] (4.1,0.5) -- (4.2,0.7);
		\node[] at (4.4,2) {$L_L$};
		\node[color = violet] at (7,2.5) {$S$};
		\draw[->-=1 rotate 0, color = red, dashed] (8.45,2.7) -- (9.2,2.7);
	\end{tikzpicture}
\caption{A Wilson or a 't Hooft loop $L$ dragged across a non-invertible defect may become an improperly quantized disc operator. The red arrow denotes the orientation of the surface.}
\label{fig:Parallel Action}
\end{figure}

In this section, we study the action of the defects, constructed in section \ref{section Maxwell defects}, on line operators. Let us begin by first considering what happens when one drags a loop operator across a defect, as depicted in figure \ref{fig:Parallel Action}. In our conventions the dual gauge field $\Tilde{A}$ is defined by (see \eqref{dualF}) 
\begin{equation}
   d \Tilde{A} = -\frac{i}{e^2} \star d A + \frac{\theta}{ 2\pi} d A\,,
\end{equation}
allowing us to write Wilson and 't Hooft loops, with integer charges $n$ and $m$ respectively, as disc operators in the following way: 
\begin{align}
    W_n(\g) &= e^{i n \oint_\g A}  = e^{i n \int_D dA} ~,\\
    H_m(\g) &= e^{-i m \oint_\g \tilde{A}}  = e^{-i m  \int_D d \tilde{A}}  = e^{- \frac{m}{e^2} \int_D \star  d A - \frac{ i m \theta }{2 \pi} \int_D dA} ~,
\end{align}
where $\partial D=\g$. We can now use the sewing conditions \eqref{Maxwell sewing} to determine the result of dragging such a loop operator across the defect from the right to the left.
The defect transforms a loop $W_n(\gamma) H_m(\gamma)$ into a disc operator
\begin{equation}
\begin{split}
 W_{n}H_{m}  \rightarrow  W_{mN+\frac{\left(n-mN_{R}\right)N_{L}}{N}}H_{\frac{mN_{R}-n}{N}}\,,
\label{loops transformation}
\end{split}
\end{equation}
where the $N$'s are given in terms of the bulk couplings $e^2$ and $\theta$ and the $SO(2)$ angle $\varphi$ as in \eqref{N in terms on couplings}, and are generically rational numbers. In the above, improperly quantized Wilson and 't Hooft loops should be viewed as disc operators.
The fact that a loop can transform into a disc operator reflects the non-invertible nature of the defect, as already discussed in \cite{Choi:2021kmx,Choi:2022zal,Choi:2022jqy}.

\begin{figure}
	\centering
	\begin{tikzpicture}[scale = 0.8]
		\node[	trapezium, 
		draw = violet, 
		minimum width=3.2cm,
		trapezium left angle=70, 
		trapezium right angle=110, 
		rotate = 20, 
		trapezium stretches=false,
		minimum height=2.4cm, 
		fill = magenta!20,
		]
		at (0.8,0.5) {};

		\draw[->-=0.7 rotate 0] (5.5,.5) -- (1,0.5);
		\draw[dashed] (-1.2,.5) to (1,0.5);
		\draw[->-=0.5 rotate 0] (-1.2,.5) -- (-4.5,0.5);
		\node[] at (4.5,1) {$L_R$};
		\node[] at (-3,1) {$L_L$};
		\node[color = violet] at (1,2.5) {$S$};
		\draw[->-=1 rotate 0, color = red, dashed] (2.45,2.7) -- (3.2,2.7);
		\node[circle,inner sep=1pt,draw, fill, color = black] at (0.9,0.5) {};
		\node[anchor = south] at (0.9,0.5) {$\phi$};
	\end{tikzpicture}
	\caption{Piercing a defect $S$ by a line operator $L$. At the junction, there is a monopole operator $\phi$ as required by gauge invariance.}
	\label{fig:S on lines}
\end{figure}
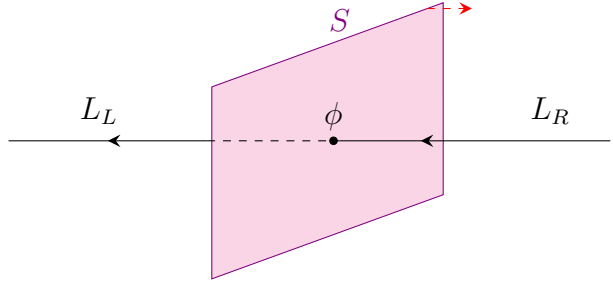

The more interesting case is when a line operator is piercing a defect, as in figure \ref{fig:S on lines}. In this case, a line with one-form symmetry charges $n_R$ and $m_R$ (see \eqref{emcharges}) that pierces the defect from the right is transformed into a line with charges $n_L$ and $m_L$ from the left. As we discuss below, using the equations of motion on the defect one can find the relation between these charges, and determine the action $S \cdot L$ of the defect $S$ on a line $L$. 

At the vertex in figure \ref{fig:S on lines}, there is a junction operator that changes the incoming line from the right to the line exiting to the left. From the three-dimensional TQFT perspective, we will see that such junctions correspond to monopole operators for the fields $A_L$ and $A_R$, as well as for other possible dynamical gauge fields living only on the defect. Monopole operators are in general not gauge invariant, and need to be attached to Wilson lines. The lines in figure \ref{fig:S on lines} are exactly what is needed in order to render the whole configuration gauge invariant. 

The action of a defect on lines can be determined by identifying the monopole operators living on it and their quantum numbers. Then, for each monopole operator, examining which combinations of lines should be attached from the two sides for gauge-invariance gives the action of defect on lines. If for an incoming line there is no monopole operator with the right quantum numbers that can transform it to an exiting line, the resulting action of the defect should be interpreted as zero. On the other hand, in some cases there may be more than one choice of monopole operators that can convert an incoming line $L_R$. In such cases, the result of the action of the defect on the line is given by the sum over all the possible such lines $L_L$. An example in which this happens is 3d Chern-Simons theory, which we review in appendix \ref{app: CS defect}.

Let us illustrate the above in the case where the defect Lagrangian is given by 
\begin{equation}
\mathcal{L}_{S}=\frac{iN_{L}}{4\pi}A_{L} \wedge dA_{L}+\frac{iN_{R}}{4\pi}A_{R} \wedge dA_{R}-\frac{iN}{2\pi}A_{L} \wedge dA_{R} \,,
\label{integer defect Lagrangian}
\end{equation}
for integer $N_L$, $N_R$, and $N$. The equations of motion on the defect for $A_L$ and $A_R$ are  
\begin{equation}
\frac{\widetilde{F}_{L}}{2\pi} {\Big |}_S=N\frac{F_{R}}{2\pi}{\Big |}_S-N_{L}\frac{F_{L}}{2\pi}{\Big |}_S\,,\qquad\frac{\widetilde{F}_{R}}{2\pi}{\Big |}_S=N_{R}\frac{F_{R}}{2\pi}{\Big |}_S-N\frac{F_{L}}{2\pi}{\Big |}_S\,.
\label{eom integer Ns}
\end{equation}
When integrated along a two-dimensional surface on the defect around the junction, we obtain the following relations between the corresponding charges: 
\begin{equation}
n_L=N m_R -N_{L} m_L\,,\qquad n_R=N_{R}m_R-N m_L \,.
\label{more general defect charges}
\end{equation}
We can solve for the left charges in terms of the right ones and get the following action:
\begin{equation}
    S \cdot W_n H_m = 
    \begin{cases}
    W_{m N + \frac{(n-mN_R)N_L}{N}} H_{\frac{mN_R - n}{N}} \quad & \text{if }N|n-m N_R \,,\\
    0 \quad &\text{otherwise}\,.
     \end{cases}
\label{action on lines naive Lagrangian}     
\end{equation}
Note that the first equation above formally coincides with \eqref{loops transformation}. However, in \eqref{loops transformation} the $N$'s take generic rational values and fractional charges are allowed, since improperly quantized loop operators are actually disc operators. Instead, in \eqref{action on lines naive Lagrangian} we consider integer $N$'s and only integer charges for the lines are allowed, as required by gauge invariance.

Let us now examine equation \eqref{action on lines naive Lagrangian} from the point of view of the monopole operators living on the defect. We have two such basic monopoles, $e^{i\phi_L}$ and $e^{i\phi_R}$, where $\phi_L$ and $\phi_R$ are the dual photons of $A_L$ and $A_R$ on the defect, respectively. Each such monopole extends in the corresponding bulk to a 't Hooft line (with the same magnetic flux). From the defect Lagrangian \eqref{integer defect Lagrangian} we see that each monopole transforms non-trivially under $U(1)_L$ and $U(1)_R$ gauge transformations. As a result, by attaching suitable Wilson lines for $A_L$ and $A_R$ to make them gauge invariant we exactly end up with \eqref{action on lines naive Lagrangian}. Let us finally comment that the monopoles $e^{i\phi_L}$ and $e^{i\phi_R}$ arise naturally when considering the dual frame in the bulk, as we explain in appendix \ref{app: S dual frame}. 

In some cases, there might be additional dynamical fields living on the defect apart from $A_L$ and $A_R$. In such cases, the 
corresponding monopole operators also have to be taken into account. As a simple example, let us consider the Lagrangian \eqref{condensation Lagrangian} describing a condensation defect (for $\kappa \neq 1$). In this case, there is also the defect gauge field $a$ and equation \eqref{charges of condensation monopole} requires that in order to get a non-zero action, the charges of the lines on the left and on the right do not  only have to be equal ($n_R=n_L$) but are also a multiple of $\kappa$. This extra requirement, coming from having a non-trivial monopole associated with $a$, leads to a non-trivial action on lines. 

To illustrate this better, let us consider another example with a defect Lagrangian of the type discussed in section \ref{general case section}. Consider the defect Lagrangian
\begin{equation}
\mathcal{L}_{S}=\mathcal{A}^{q,1}\left[\frac{rdA_{R}+ldA_{L}}{q}\right]-\frac{ik}{2\pi}A_{L} \wedge dA_{R}=\frac{iq}{4\pi}a \wedge da+\frac{i}{2\pi}a \wedge \left(rdA_{R}+ldA_{L}\right)-\frac{ik}{2\pi}A_{L} \wedge dA_{R} \,,
\label{Curly A monopole example}
\end{equation}
where $a$ is a dynamical $U(1)$ gauge field living on the defect and $k$ is an integer. The equations of motion for $A_L$, $A_R$ and $a$ on the defect yield the following relations between the charges 
\begin{equation}
n_{L}=-lm_{a}+km_{R}\,,\qquad n_{R}=rm_{a}-km_{L}\,,\qquad qm_{a}+rm_{R}+lm_{L}=0 \,,
\end{equation}
where $m_a$ is the magnetic flux of the monopole of $a$. Solving as before for the left charges in terms of the right ones, we obtain the following action of the defect on lines: 
\begin{equation}
    S\cdot W_{n}H_{m}=
    \begin{cases}
    W_{mN+\frac{(n-mN_{R})N_{L}}{N}}H_{\frac{mN_{R}-n}{N}}\qquad & \text{if }lr+kq\,|\,ln-krm\text{ and }qn+r^{2}m\,,\\
    0\qquad & \text{otherwise}\,,
    \end{cases}
\label{action on lines curly A}
\end{equation}
where $N_L$, $N_R$ and $N$ are given by
\begin{equation}
    N_L =-\frac{p}{q} l^2 \,, \qquad
    N_R =-\frac{p}{q} r^2 \,, \qquad
    N = k  +\frac{p}{q} l r \,.
\end{equation}
Notice that the condition for having a non-zero defect action written in the first line of \eqref{action on lines curly A} takes into account the requirement that the monopole flux $m_a$ is an integer,\footnote{Indeed, in this case $m_{a}=\frac{ln-krm}{lr+kq}$ .} which (as in the case of the condensation defect) adds an extra restriction compared to only demanding that the charges of the left lines are integer. Indeed, the relation between the left and right charges in \eqref{action on lines curly A} corresponds to the sewing conditions \eqref{Maxwell sewing}, however using them alone will miss those lines that are in fact annihilated by the defect. As an example, let us consider the defect with $q=8$, $l=4$, $r=2$, $N=1$ and its action on the line $W_3H_2$. Naively, using \eqref{action on lines curly A} while ignoring the specified conditions on the right-hand side (or using only the sewing conditions) it might seem consistent that the result of the defect action is given by the left line $H_2$. However, in this case $m_a=1/2$ and this combination of lines is annihilated by the defect.

In the case of a generic defect of the type discussed in section \ref{general case section} (see equation \eqref{generic Ls}), the action on lines is given by (see appendix \ref{app: generic defects} for details) 
\begin{equation}
S\cdot W_{n}H_{m}=\begin{cases}
W_{mN+\frac{(n-mN_{R})N_{L}}{N}}H_{\frac{mN_{R}-n}{N}}\qquad & \text{if }d\,|\,w,\,h,\,u_{1},\text{ and }u_{2}\,,\\
0\qquad & \text{otherwise}\,,
\end{cases}
\label{action on lines generic defect}
\end{equation}
where $N_L$, $N_R$ and $N$ are given in \eqref{Ns in terms of plrn} and are generically rational, and 
\begin{equation}
\begin{split}
d&=k q_{1}q_{2}+l_{1}p_{1}q_{2}r_{1}+l_{2}p_{2}q_{1}r_{2}\,,\\
u_{i}&=l_{i}nq_{i+1}-l_{i}mk_{R}q_{i+1}-mkq_{i+1}r_{i}-l_{i+1}mp_{i+1}r_{1}r_{2}+l_{i}mp_{i+1}r_{i+1}^{2}\,,\\
h&=-nq_{1}q_{2}+mk_{R}q_{1}q_{2}-mp_{1}q_{2}r_{1}^{2}-mp_{2}q_{1}r_{2}^{2}\,,\\
w&=q_{1}q_{2}\left(nk_{L}+mk^{2}-mk_{L}k_{R}\right)+p_{2}q_{1}\left(2ml_{2}kr_{2}+mk_{L}r_{2}^{2}-l_{2}^{2}n+l_{2}^{2}mk_{R}\right)\\&+p_{1}q_{2}\left(l_{1}^{2}mk_{R}+2l_{1}mkr_{1}+mk_{L}r_{1}^{2}-l_{1}^{2}n\right)-mp_{1}p_{2}\left(l_{1}r_{2}-l_{2}r_{1}\right)^2\,,
\end{split}
\label{action on lines conditions generic}
\end{equation}
where the subscript $i=1,2$ should be regarded modulo 2. As expected, we note that the result of equation \eqref{action on lines generic defect} is the same as in \eqref{action on lines naive Lagrangian}, but for the (more restrictive) conditions on the various integer coefficients for having a non-zero defect action.

\section{Topological defects in 2d scalar theory}
\label{section scalar defects}

In this section, we repeat our discussion in the case of the 2d compact scalar $\phi \sim \phi + 2\pi$ at radius $R$. The topological defects we consider in this section have been previously studied in various works \cite{Fuchs:2007tx,Kapustin:2009av,Thorngren:2021yso,Choi:2022zal,Choi:2021kmx}, and our goal here is to parallel our discussion of Maxwell theory and see how the 2d compact scalar fits in this picture. 

The Euclidean Lagrangian is given by 
\begin{equation}
\LL(R;\phi) = \frac{R^2}{4\pi} d\phi \wedge \star d\phi \,.
\end{equation}
The theory has the discrete $\mathbb{Z}_2$ zero-form symmetry $\f \rightarrow -\f$ and two continuous $U(1)$ zero-form symmetries associated to constant shifts of the scalar and its dual. The corresponding charges are
\begin{equation}
m = \int_{\g} \frac{d\tilde{\phi}}{2\pi}\,,
\qquad w = \int_{\g} \frac{d\phi}{2\pi} \,,
\label{scalarcharges}
\end{equation}
where the dual scalar is defined by
\begin{equation}
    d\tilde{\f} = - i R^2 \star d\f\,.
\end{equation}
We refer to them as $U(1)_m$ and $U(1)_w$, where the subscripts stand for momentum and winding.

This theory also enjoys $\T$-duality which exchanges $d\phi$ and $d\tilde{\phi}$, while acting on the radius as $R\rightarrow 1/R$. At the self-dual point $R=1$, it becomes a global symmetry exchanging $d\phi$ and $i \star d\f$. In the rest of this section, we refer to a defect implementing this transformation (while leaving the radius invariant) as $\T$ defect. 

As in the Maxwell case, we define a codimension-one defect (i.e.\,a line) by the action
\begin{equation}
S=\int_{S^-} \LL (R;\phi_L) + \int_{S^+} \LL (R;\phi_R) + \int_{S} \LL_S (\phi_L,\phi_R,b) \,.
\label{actionscalar}
\end{equation}
Our goal is to investigate defects that mix $d\f$ and $i\star d\f$ as
\begin{align}
    d\phi_L|_S &= \alpha \, d\phi_R|_S + \beta \; i \star d\phi_R|_S\,, \label{transformation3}\\
    i \star d\phi_L|_S &= \gamma \, d\phi_R|_S + \delta \; i \star d\phi_R|_S \,,
\label{transformation4}
\end{align}
where $\alpha$, $\beta$, $\gamma$ and $\delta$ are real parameters. As in the Maxwell case, in order to have a topological defect we need to impose the condition \eqref{matchingtensor}, where now the energy-momentum tensor is given by
\begin{equation}
T_{\mu\nu}(\phi) = \frac{R^2}{2\pi}\left(\partial_\mu\phi \, \partial_\nu\phi - \frac{1}{2} g_{\mu\nu} \partial_\alpha \phi \, \partial^\alpha \phi\right) \,.
\label{tensorscalar}
\end{equation}
In contrast with the Maxwell theory, in this case we only get four discrete solutions. They are as follows:

\begin{itemize}
\item The trivial defect ($d\phi_L|_S=d\phi_R|_S$, $\star d\phi_L|_S=\star d\phi_R|_S$) for an arbitrary value of $R^2$,
\begin{equation}
\mathbbm{1}: \quad  \frac{i}{2\pi} \varphi (d\phi_L - d\phi_R)~,
\label{trivial}
\end{equation}
where $\varphi$ is a compact scalar field living on the defect. We can also multiply this defect Lagrangian by an integer $\kappa$, as in the case of the condensation defect in Maxwell theory (see equation \eqref{condensation Lagrangian}).
However, in this case, the defect becomes non-simple and it is rather a direct sum of $U(1)_m$ defects (see appendix \ref{app: T fusion}). This is an example where naively choosing a defect Lagrangian might a priori look like giving rise to a non-invertible defect, but it actually turns out to be a direct sum of invertible ones.  
\item The $\mathbb{Z}_2$ defect ($d\phi_L|_S=-d\phi_R|_S$, $\star d\phi_L|_S=-\star d\phi_R|_S$) for an arbitrary value of $R^2$,
\begin{equation}
\mathbb{Z}_2: \quad \frac{i}{2\pi} \varphi (d\phi_L + d\phi_R)~,
\label{trivial and Z2}
\end{equation}
where $\varphi$ again is a compact scalar field living on the defect. Similarly to the trivial defect, multiplying the above Lagrangian by an integer makes the defect non-simple. Note that it is straightforward to generalize the Lagrangian construction of the trivial and the $\mathbb{Z}_2$ defects for a free scalar theory in arbitrary dimension.
\item The $\T$ defect ($d\phi_L|_S=i\star d\phi_R|_S$)  which is described by the Lagrangian 
\begin{equation}
\T: \quad \frac{i N_1}{2\pi} \phi_L  d\varphi_1 +\frac{i N_2}{2\pi} \phi_R  d\varphi_2 - \frac{i N_{12}}{2\pi} \varphi_1   d\varphi_2 \,,
\label{tdualitylagrangian}
\end{equation}
where $\varphi_1$ and $\varphi_2$ are two compact scalars living on the defect, and the $N_i$'s are integers. This realizes the $\T$ defect for rational values of the radius squared:
\begin{equation}
R^2 = \frac{N_1 N_2}{N_{12}}\in \mathbb{Q}\,.
\end{equation}
We see that there are different values of the $N_i$'s leading to the same radius. To avoid overcounting defects we can just make the choice $N_1=1$, with $N_2$ and $N_{12}$ coprime. The case with $R^2\in \mathbb{N}$ was previously considered in \cite{Choi:2021kmx}, and corresponds to the choice $N_{12}=1$ which leads to the defect Lagrangian 
\begin{equation}
\frac{i }{2\pi} \phi_L  d\varphi_1 +\frac{i N_2}{2\pi} \phi_R  d\varphi_2 - \frac{i }{2\pi} \varphi_1   d\varphi_2 \,.
\end{equation}
Performing the change of variables $\varphi_1 \rightarrow \varphi_1 + N_2 \f_R$ and $\varphi_2 \rightarrow \varphi_2 - \f_L$ we end up with
\begin{equation}
\frac{iN_2}{2\pi} \phi_L  d\f_R - \frac{i }{2\pi} \varphi_1   d\varphi_2 \,,
\end{equation}
which is the same defect Lagrangian considered in \cite{Choi:2021kmx} plus a trivial TQFT.

 Let us stress again that this defect exchanges $d\phi$ and $i\star d\phi$, but leaves the bulk theory invariant (and hence $R$ is the same on both sides).
In appendix \ref{app: T fusion} we compute the fusion of two $\T$ defects in the case of integer $R^2=N$. For $N=1$ it is invertible, while for $N \neq 1$ we show that $\T\times \T$ is a sum of $U(1)_m$ defects \cite{Fuchs:2007tx}.
\item Finally, we also have the product $\mathbb{Z}_2 \times \T$ defect ($d\phi_L|_S=-i\star d\phi_R|_S$) for $R^2 \in \mathbb{Q}$.
\end{itemize}
\pagebreak
\section*{Acknowledgements}
We would like to thank Thomas Dumitrescu for multiple discussions. We are grateful to Lorenzo Di Pietro, Luigi Tizzano, Sahand Seifnashri, and Shu-Heng Shao for comments on the manuscript. OS would like to thank Gabi Zafrir, and PN would like to thank Edoardo Lauria and Jeremias Aguilera Damia for related discussions. Our work is supported by the Mani L.~Bhaumik Institute for Theoretical Physics. PN is supported by a DOE Early Career Award under DE-SC0020421.

\appendix

\section{Defects from gauging and dualities}
\label{app: Gauging}

In this appendix we show how to construct defects of the type considered in section \ref{general case section} by combining gauging of discrete subgroups of the one-form symmetry with elements of the $SL(2,\mathbbm{Z})$ duality, in the spirit of the recent literature \cite{Choi:2021kmx,Choi:2022jqy,Choi:2022rfe,Choi:2022zal}.
As we discussed in section \ref{section action on lines}, such defects act on the one-form electric and magnetic charges as 
\begin{equation}
    \begin{pmatrix}
n \\
m
\end{pmatrix}
\rightarrow 
    \begin{pmatrix}
\frac{N_L}{N} & N-\frac{N_L N_R}{N} \\
-\frac{1}{N}  & \frac{N_R}{N} 
\end{pmatrix}
    \begin{pmatrix}
n  \\
m 
\end{pmatrix}\,,
\label{charges transformation}
\end{equation}
where the $N$'s are generally rational. Notice that the above matrix has unit determinant and therefore it is an element of $SL(2,\mathbbm{Q})$. In the following, we show that such transformation can be also obtained by a sequence of $SL(2,\mathbbm{Z})$ duality transformations along with gauging of discrete subgroups of the one-form symmetry. 

We can decompose the above matrix as 
\begin{equation}
U \cdot G \cdot T =     \begin{pmatrix}
1 & 0 \\
y  & 1
\end{pmatrix}
\begin{pmatrix}
z  & 0 \\
0 & z^{-1}
\end{pmatrix}
\begin{pmatrix}
1  & x \\
0 & 1
\end{pmatrix}
~,
\end{equation}
where
\begin{equation}
    x = \frac{N^2}{N_L}-N_R \equiv \frac{p_x}{q_x}\,, \qquad y = -\frac{1}{N_L} \equiv \frac{p_y}{q_y}\,, \qquad z = \frac{N_L}{N} \equiv \frac{p_z}{q_z}~,
\end{equation}
with $p_i,q_i\in\mathbbm{Z}$ such that gcd$(p_i,q_i)=1$. A $G$-transformation is just the gauging of a $\mathbbm{Z}_{q}^{e}\times \mathbbm{Z}_{p}^m$ subgroup of the $U(1)_e\times U(1)_m$ one-form symmetry. To see this, consider for simplicity only the gauging of a $\mathbbm{Z}_{q}^{e}$ electric one-form symmetry.
After the gauging we have a $U(1)/\mathbbm{Z}_q$ gauge theory with the same coupling $\tau$, but whose gauge field $\tilde{A}$ has fluxes that are multiples of $\frac1q$. This $U(1)/\mathbbm{Z}_q$ theory can then be rewritten as a $U(1)$ theory by redefining $A'=q \tilde{A}$, thus changing the coupling to $\t'= \t/q^2$.
In this procedure, due to the rescaling, note that the line with charges $(q,0)$ in the original theory before gauging maps to the line with charges $(1,0)$. We therefore see that gauging a $\mathbbm{Z}_{q}^{e}$ subgroup of the electric one-form symmetry provides a map from a $U(1)$ gauge theory with coupling $\t$ to another one with coupling $\t/q^2$, and the charges map as 
\begin{equation}
    \begin{pmatrix}
n \\
m
\end{pmatrix}
\rightarrow 
    \begin{pmatrix}
\frac{1}{q} & 0 \\
0  & q
\end{pmatrix}
    \begin{pmatrix}
n  \\
m 
\end{pmatrix}\,.
\end{equation}
Similarly, the gauging of a $\mathbbm{Z}_{p}^m$ magnetic one-form symmetry results in the rescaling $\t \rightarrow p^2 \tau$ and the map on charges
\begin{equation}
    \begin{pmatrix}
n \\
m
\end{pmatrix}
\rightarrow 
    \begin{pmatrix}
p & 0 \\
0  & \frac{1}{p}
\end{pmatrix}
    \begin{pmatrix}
n  \\
m 
\end{pmatrix}\,.
\end{equation}

A $T$-transformation is a combination of two $G$-transformations and a $\T$-duality,\footnote{A $T$-transformation can also be obtained by stacking SPT phases and gauging a $\mathbbm{Z}_q^m\in U(1)_m$ magnetic one-from symmetry, an operation usually referred to as $CT^{(k)}S^{(k)}T^{(k)}$ gauging (see  e.g. \cite{Choi:2022jqy}).} as it can be seen from 
\begin{equation}
    \begin{pmatrix}
1  & \frac{p}{q} \\
0 & 1
\end{pmatrix}
=
\begin{pmatrix}
\frac{1}{q}  & 0 \\
0 & q
\end{pmatrix}
\begin{pmatrix}
1  & pq \\
0 & 1
\end{pmatrix}
\begin{pmatrix}
q  & 0 \\
0 & \frac{1}{q}
\end{pmatrix}
\,,
\end{equation}
while a $U$-transformation can be obtained from $T$ using $\cS$-duality as follows
\begin{equation}
    \begin{pmatrix}
1  & 0 \\
y & 1
\end{pmatrix}
=
\begin{pmatrix}
0  & 1 \\
-1 & 0
\end{pmatrix}
\begin{pmatrix}
1  & -y \\
0 & 1
\end{pmatrix}
\begin{pmatrix}
0  & -1 \\
1 & 0
\end{pmatrix}
\,.
\end{equation}
Most importantly, we see that \eqref{charges transformation} can be obtained from a series of transformations which leave the coupling $\tau$ invariant, since
\begin{equation}
\tau \quad \rightarrow \quad
\frac{ \frac{N_L}{N}\tau+N-\frac{N_L N_R}{N}}{-\frac{1}{N}\tau+\frac{N_R}{N}} = \tau~,
\end{equation}
if
\begin{equation}
\tau = \frac{\theta}{2\pi}+\frac{i}{e^2} = \frac{N_R-N_L}{2} + i \sqrt{N^2 - \left(\frac{N_R + N_L}{2}\right)^2}~,
\end{equation}
as in equation \eqref{constraint2}.
Hence, we conclude that one can construct defects that generate an $SO(2)$ rotation on $F$ and $\star F$ with $\cos\varphi\in \mathbbm{Q}$, by combining the gauging of discrete subgroups of the one-form symmetry on half-space with elements of the $SL(2,\mathbbm{Z})$ duality group.

\section{Action of surface defects in Chern-Simons theory}
\label{app: CS defect}

In this appendix, we review the action of condensation defects on lines in Chern-Simons theory and we study their action on line operators in the spirit of section \ref{section action on lines}. We consider their Lagrangian description as in \cite{Roumpedakis:2022aik} and we analyze the scalar operators which serve as junctions between Wilson lines on the left and right of the defect. These operators give a nice interpretation for the action of the defect on lines. In particular, if there are multiple allowed junction operators then the action of the defect on lines is not simple and results in a sum of several terms.  

In \cite{Roumpedakis:2022aik}, it was shown that topological surfaces in $U(1)_{2N}$ Chern-Simons theory are labeled by a divisor of $N=nm$ and can be described by the action
\begin{equation}
S_{CS} = {2iN\over 4\pi } \int_{S^-} A_L \wedge dA_L 
+{2iN\over 4\pi} \int_{S^+} A_R \wedge dA_R 
+ \int_S \left( {i  n\over 2\pi} d\phi \wedge (A_L- A_R) +{iN\over 2\pi} \int_{S} A_L \wedge  A_R \right),
\label{CS}
\end{equation}
which is invariant under the gauge transformations
\begin{equation}
A_L  \rightarrow A_L + d\a_L\,, \qquad A_R  \rightarrow A_R + d\a_R\,,\qquad \f   \rightarrow \f + m(\a_L + \a_R) \,.
\end{equation}
The equations of motion on the defect are
\begin{equation}
dA_L|_S = dA_R|_S \,, \qquad d\f = m(A_L + A_R)|_S \,.
\label{CS eom 1}
\end{equation}
Considering the second equation, the line integral of the left-hand side around the junction measures the winding of $\f$
\begin{equation}
\oint_\g \frac{d \f}{2\p} = \l \in \mathbb{Z} \,,
\label{lambda winding}
\end{equation}
while the integral of the right-hand side measures the one-form charges of left and right Wilson lines, leading to
\begin{equation}
\l=m\left(\frac{q_{L}}{2N}+\frac{q_{R}}{2N}\right)~.
\end{equation}
Using the fact that $N=nm$ we arrive at 
\begin{equation}
q_{L}+q_{R}=2n\l~,
\end{equation}
where $\l=0,\dots,m-1$ since the charges in the bulk are defined modulo $2N=2nm$. For each $\l$ we have an operator sitting at the junction which converts a line coming from the right to a line exiting to the left of the defect. We can then write the action on lines as 
\begin{equation}
S_n \cdot W_q = \sum_{\substack{\l=0 \\  q - n \l=0~\text{mod}~m }}^{m-1} W_{2n\l-q} \,.
\label{action on lines CS new}
\end{equation}
This is in agreement with the result of \cite{Roumpedakis:2022aik} (see equation 6.16), where it was shown that the action of a defect on a Wilson line of charge $q$ is given by 
\begin{equation}
S_n \cdot W_q = \sum_{\substack{k=0 \\ mk +q=0~\text{mod}~n }}^{n-1} W_{q+2mk} \,.
\label{action on lines CS}
\end{equation}

\begin{figure}
	\centering
	\begin{tikzpicture}[scale = 0.8]
		\node[	trapezium, 
		draw = violet, 
		minimum width=3.2cm,
		trapezium left angle=70, 
		trapezium right angle=110, 
		rotate = 20, 
		trapezium stretches=false,
		minimum height=2.4cm, 
		fill = magenta!20,
		]
		at (0.8,0.5) {};

		\draw[->-=0.7 rotate 0] (5.5,.5) -- (1,0.5);
		\draw[dashed] (-1.2,.5) to (1,0.5);
		\draw[->-=0.5 rotate 0] (-1.2,.5) -- (-4.5,0.5);
		\node[] at (4.5,1) {$W_R = e^{i q_R \int A_R}$};
		\node[] at (-3,1) {$W_L = e^{i q_L \int A_L}$};
		\node[color = violet] at (1,2.5) {$S$};
		\draw[->-=1 rotate 0, color = red, dashed] (2.45,2.7) -- (3.2,2.7);
		\node[circle,inner sep=1pt,draw, fill, color = black] at (0.9,0.5) {};
		\node[anchor = south] at (0.9,0.5) {$e^{ik \phi + i\l \bar\phi}$};
	\end{tikzpicture}
	\caption{Gauge-invariant configuration of Wilson lines and twist operators on a defect $S$ in $U(1)_{2N}$ Chern-Simons theory.}
	\label{fig:action on W}
\end{figure}
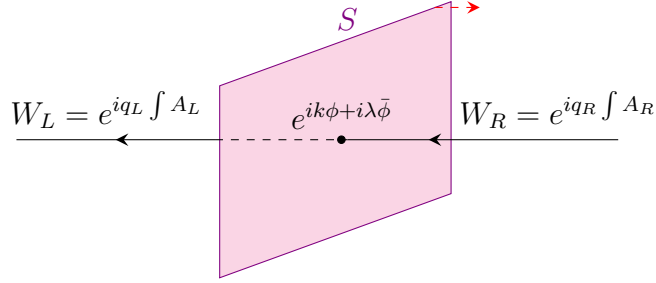

An equivalent way to determine the action on lines is by considering the scalar operators on the defect that can sit at the junction.
Apart from the operator $e^{i k \phi}$ there is also another scalar living on the defect, namely the twist operator $e^{i \l \bar{\f}}$ around which $\phi$ has winding $\l$, as in \eqref{lambda winding}.\footnote{One way to analyze this operator is by dualizing $\f$ in \eqref{CS}. The Lagrangian becomes \begin{equation}
\int_{S}\left(
{i n\over 2\pi} b \wedge (A_L-A_R) - {i\over 2\pi} \bar\phi \, db +{iN\over 2\pi} A_L \wedge A_R\right)\,,
\end{equation}
and the additional gauge transformation for $\bar\phi$ reads $\bar{\f} \rightarrow \bar\f + n(\a_L - \a_R)$.
}
Because of the third term in \eqref{CS} the twist operator $e^{i \l \bar{\f}}$ is not gauge invariant and should be attached to a Wilson line of $A_L$ of charge $n$, and to a Wilson line of $A_R$ of the same charge. In order to transform a Wilson line $W_R$ coming from the right into a Wilson line $W_L$ exiting to the left in a gauge-invariant way, we need a gauge invariant junction on $S$. If we consider the operator $\exp(ik\phi+i\l\bar\phi)$ at the junction (see figure \ref{fig:action on W}),
gauge invariance requires 
\begin{equation}
    q_L - k m -\l n = 0 \,, \qquad q_R + k m - \l n =0 \,.
    \label{charges in CS}
\end{equation}
This implies 
\begin{equation}
    q_L = q_R + 2mk\,,
    \label{LR charges in CS}
\end{equation}
where $k=0,1, \dots , n-1$ labels the ways to convert a right Wilson line into a left Wilson line. This leads to \eqref{action on lines CS} or, equivalently, to \eqref{action on lines CS new} if in \eqref{charges in CS} we eliminate $k$ instead of $\l$.

\section{$\mathcal{S}$ defect at integer coupling}
\label{app: S dual frame}

Let us examine the action on lines of the $\cS$ defect constructed in \cite{Choi:2021kmx}, using the dual-frame description of one of the bulk sides.
We consider the action at $\theta=0$ and $1/e^2=N \in \mathbb{N}$, which reads\footnote{Notice that there is a sign difference in the defect Lagrangian in \eqref{S-duality defect action} compared to the one in \cite{Choi:2021kmx}.} 
\begin{equation}
S=\frac{N}{4\pi}\int_{S^{-}}dA_{L}\wedge\star dA_{L}-\frac{iN}{2\pi}\int_{S}A_{L}\wedge dA_{R}+\frac{N}{4\pi}\int_{S^{+}}dA_{R}\wedge\star dA_{R} \,,
\label{S-duality defect action}
\end{equation}
and choose to dualize one of the sides, say the right one. Following \cite{Witten:1995gf}, we rewrite \eqref{S-duality defect action} as 
\begin{equation}
\begin{split}
    S =&  \frac{N}{4\pi}\int_{S^-} dA_L\wedge \star dA_L - \frac{iN}{2\pi} \int_{S}A_L \wedge (dA_R-G_R) -\frac{i}{2\pi} \int_{S}\phi_R dG_R \\
    & +\frac{N}{4\pi}\int_{S^+} (dA_R-G_R)\wedge \star (dA_R-G_R)-\frac{i}{2\pi}\int_{S^+} V_R\wedge dG_R \,,
\end{split}
\label{simplest S duality}
\end{equation}
where $V_R$ and $G_R$ are one-form and two-form fields, respectively. In the presence of the defect, we are also forced to introduce a scalar field $\phi_R$ living on the defect in order to preserve the gauge transformations
\begin{equation}
\begin{split}
    A_{L}\rightarrow A_{L}+d\alpha_{L}\;,\qquad &A_{R}\rightarrow A_{R}+b_{R}\;,\qquad G_{R}\rightarrow G_{R}+db_{R} \,,\\ 
    V_{R}\rightarrow V_{R}+d\widetilde{\alpha}_{R}\;&,\qquad\phi_{R}\rightarrow\phi_{R}-N\alpha_{L}+\widetilde{\alpha}_{R}\,.
\end{split}
\label{gauge trns dual}
\end{equation}
This scalar is the dual photon of $A_R$ on the defect, and $e^{i\phi_R}$ is the corresponding monopole operator. Notice that this monopole transforms non-trivially also under the gauge group $U(1)_L$, as is evident both from \eqref{S-duality defect action} and \eqref{gauge trns dual}. We can now gauge fix $A_R$ to zero, and arrive at 
\begin{equation}
\begin{split}
    S =&  \frac{N}{4\pi}\int_{S^-} dA_L\wedge \star dA_L + \frac{iN}{2\pi} \int_{S}A_L \wedge G_R -\frac{i}{2\pi} \int_{S}\phi_R  dG_R \\
    & +\frac{N}{4\pi}\int_{S^+} G_R\wedge \star G_R-\frac{i}{2\pi}\int_{S^+} V_R\wedge dG_R\,.
\end{split}
\end{equation}
The equation of motion for $G_R$ on the defect is  
\begin{equation}
     V_R|_S = N A_L|_S + d\phi_R\,.
\end{equation}
Using this equation, we can see that a Wilson line for $V_R$ (or equivalently a 't Hooft line for $A_R$) with charge $p$ (for $p\in\mathbb{Z}$) coming from the right of the defect is transformed into a Wilson line with charge $pN$ from the left, matching \eqref{action on lines naive Lagrangian}. 

Let us examine in more detail what this configuration looks like. We begin with a 't Hooft line $\exp(ip\int V_{R})$ coming from the right and ending on the defect. Under the gauge transformations \eqref{gauge trns dual} it transforms by a phase $\exp(ip\widetilde{\alpha}_{R})$, which can in turn be canceled by inserting the monopole operator $\exp(-ip\phi_R)$ at its end. However, this configuration is still not gauge invariant and transforms by the phase $\exp(ipN\alpha_{L})$ which is then canceled by attaching the Wilson line $\exp(ipN\int A_{L})$ exiting the defect to the left. The same is true also in the other direction, and a Wilson line with charge $pN$ piercing the defect from the left is transformed into a 't Hooft line with charge $p$ from the right (where now the monopole at the junction is $\exp(ip\phi_R)$). From this description, it is clear that if the charge of this Wilson line is not a multiple of $N$, it cannot pierce the defect in a gauge invariant way and is therefore annihilated by it. Notice also that a similar dualization can be done to the left side, reproducing the full picture in \eqref{action on lines naive Lagrangian} which in this case is given by 
\begin{equation}
    \mathcal{S}\cdot W_{n}H_{m}=\begin{cases}
W_{mN}H_{-n/N}\quad & \text{if }N|n\,,\\
0\quad & \text{otherwise}\,.
\end{cases}
    \label{S duality action}
\end{equation}

\section{Action of generic defects on lines}

\label{app: generic defects}
In this appendix, we derive the result \eqref{action on lines generic defect} for the action of a generic defect on lines. Consider the full action
\begin{equation}
S=\int_{S^{-}}\left(\frac{1}{4\pi e^{2}}F_{L}\wedge\star F_{L}+\frac{i\theta}{8\pi^{2}}F_{L}\wedge F_{L}\right)+\int_{S^{+}}\left(\frac{1}{4\pi e^{2}}F_{R}\wedge\star F_{R}+\frac{i\theta}{8\pi^{2}}F_{R}\wedge F_{R}\right)+\int_{S}\mathcal{L}_{S}(A_L,A_R) \,,
\label{Maxwell bulk action}
\end{equation}
where $\mathcal{L}_{S}$ is the generic defect Lagrangian \eqref{generic Ls}, explicitly given by 
\begin{equation}
\begin{split}
\mathcal{L}_{S}=&\frac{ik_{L}}{4\pi}A_{L} \wedge dA_{L}+\frac{ik_{R}}{4\pi}A_{R} \wedge dA_{R}-\frac{ik}{2\pi}A_{L} \wedge dA_{R}\\
&+\mathcal{A}^{q_{1},p_{1}}\left[\frac{l_{1}\,dA_{L}+r_{1}\,dA_{R}}{q_{1}}\right]+\mathcal{A}^{q_{2},p_{2}}\left[\frac{l_{2}\,dA_{L}+r_{2}\,dA_{R}}{q_{2}}\right].
\end{split}
\end{equation}

For generic values of the parameters, there is no simple description for the above defect Lagrangian. Nevertheless, the $\A^{q,p}$ TQFT can be realized on the boundary of a $3+1$ dimensional model. In the following, we  make use of the folding trick and rewrite the bulk theory \eqref{Maxwell bulk action} as a theory only on the right of the defect $S^+$ with a boundary on $S$. We then use the aforementioned $3+1$ dimensional description of the defect TQFT. For simplicity, we consider a flat boundary at $x=0$. 

We can replace the left side of the defect by its $x$-reversal living on $S^+$. Then, the bulk action is given by 
\begin{equation}
S=\int_{S^{+}}\left(\frac{1}{4\pi e^{2}}F_{R}\wedge\star F_{R}+\frac{i\theta}{8\pi^{2}}F_{R}\wedge F_{R}+\frac{1}{4\pi e^{2}}\overline{F}_{L}\wedge\star \overline{F}_{L}-\frac{i\theta}{8\pi^{2}}\overline{F}_{L}\wedge \overline{F}_{L}\right)+\int_{S}\mathcal{L}_{S}(A_L,A_R)\,.
\label{action on S+}
\end{equation}
where $\overline{F} = d \overline{A}$ satisfies
\begin{equation}
    \overline{F}|_S = F|_S, \quad \star \overline{F}|_S = -\star F|_S~.
\end{equation}
The minimal $\A^{q,p}$ TQFT can be realized on the boundary of a  $3+1$ dimensional system \cite{Hsin:2018vcg} with action
\begin{equation}
\int_{S^{+}}\left(\frac{iq}{2\pi}b\land dc+\frac{iqh}{4\pi}b\land b+\frac{i}{2\pi}b\land dA\right) \,,
\end{equation}
where $b$ is a two-form gauge field with Dirichlet boundary conditions on $S$, $c$ is a one-form gauge field constraining $b$ to be a $\mathbb{Z}_q$ gauge field, and $h$ is the inverse of $p$ mod $q$. The above is equivalent to the combined action 
\begin{equation}
-\int_{S^{+}}\frac{ip}{4\pi q} dA\land dA+\int_{S}\mathcal{A}^{q,p}\left[dA/q\right] \,.
\end{equation}
Hence, we can replace \eqref{Maxwell bulk action} by
\begin{multline}
    S=\int_{S^{+}}\left(\frac{1}{4\pi e^{2}}F_{R}\wedge\star F_{R}+\frac{i}{4\pi}\left(\frac{\theta}{2\pi}-N_{R}\right)F_{R}\wedge F_{R}
    +\frac{1}{4\pi e^{2}}  \overline{F}_{L}\wedge\star  \overline{F}_{L}
    -\frac{i}{4\pi}\left(\frac{\theta}{2\pi}+N_{L}\right) \overline{F}_{L}\wedge  \overline{F}_{L}\right)\\
    +\int_{S^{+}}\frac{iN}{2\pi} \overline{F}_{L}\wedge F_{R}+ \int_{S^{+}}\left(\frac{iq_{1}}{2\pi}b_{1}\land dc_{1}+\frac{iq_{1}h_{1}}{4\pi}b_{1}\land b_{1}+\frac{i}{2\pi}b_{1}\land\left(l_{1}\overline {F}_{L}+r_{1}F_{R}\right)\right)\\
    +\int_{S^{+}}\left(\frac{iq_{2}}{2\pi}b_{2}\land dc_{2}+\frac{iq_{2}h_{2}}{4\pi}b_{2}\land b_{2}+\frac{i}{2\pi}b_{2}\land\left(l_{2}\overline {F}_{L}+r_{2}F_{R}\right)\right)\,,
\end{multline}
where $N_L$, $N_R$, and $N$ are defined as in \eqref{Ns in terms of plrn}. Now, the equations of motion for $A_{L}$, $A_{R}$, and $b_{i}$ on the boundary $S$ (where $b_{i}=0$ due to the Dirichlet boundary conditions) are, respectively 
\begin{equation}
\frac{\widetilde{F}_{L}}{2\pi}{\Big |}_S+N_{L}\frac{F_{L}}{2\pi}{\Big |}_S-N\frac{F_{R}}{2\pi}{\Big |}_S=0\,,\qquad
\frac{\widetilde{F}_{R}}{2\pi}{\Big |}_S+N\frac{F_{L}}{2\pi}{\Big |}_S-N_{R}\frac{F_{R}}{2\pi}{\Big |}_S=0 \,,
\end{equation}
and 
\begin{equation}
q_{1}\frac{dc_{1}}{2\pi}{\Big |}_S+l_{1}\frac{F_{L}}{2\pi}{\Big |}_S+r_{1}\frac{F_{R}}{2\pi}{\Big |}_S=0\,,\qquad
q_{2}\frac{dc_{2}}{2\pi}{\Big |}_S+l_{2}\frac{F_{L}}{2\pi}{\Big |}_S+r_{2}\frac{F_{R}}{2\pi}{\Big |}_S=0 \,.
\label{extra conditions - fluxes}
\end{equation}
The first two equations are analogous to the case with integer $N$'s given in \eqref{eom integer Ns}. They are just a consequence of the sewing conditions, and imply 
\begin{equation}
n_{L}=Nm_{R}-N_{L}m_{L}\,,\qquad n_{R}=N_{R}m_{R}-Nm_{L}\,.
\label{relations for generic charges}
\end{equation}
In addition, the equations in \eqref{extra conditions - fluxes} give extra conditions securing integer monopole fluxes $m_{c_{1}}$ and $m_{c_{2}}$  
\begin{equation}
q_{1}m_{c_{1}}+l_{1}m_{L}+r_{1}m_{R}=0\,,\qquad q_{2}m_{c_{2}}+l_{2}m_{L}+r_{2}m_{R}=0\,.
\label{generic extra conditions}
\end{equation}
Overall, combining equations \eqref{relations for generic charges} and \eqref{generic extra conditions} we obtain the following action of a generic defect of the type discussed in section \ref{general case section} on lines:  
\begin{equation}
S\cdot W_{n}H_{m}=\begin{cases}
W_{mN+\frac{(n-mN_{R})N_{L}}{N}}H_{\frac{mN_{R}-n}{N}}\qquad & \text{if }d\,|\,w,\,h,\,u_{1}\text{ and }u_{2}\,,\\
0\qquad & \text{otherwise} \,,
\end{cases}
\end{equation}
where  
\begin{equation}
\begin{split}
d&=k q_{1}q_{2}+l_{1}p_{1}q_{2}r_{1}+l_{2}p_{2}q_{1}r_{2}\,,\\
u_{i}&=l_{i}nq_{i+1}-l_{i}mk_{R}q_{i+1}-mkq_{i+1}r_{i}-l_{i+1}mp_{i+1}r_{1}r_{2}+l_{i}mp_{i+1}r_{i+1}^{2}\,,\\
h&=-nq_{1}q_{2}+mk_{R}q_{1}q_{2}-mp_{1}q_{2}r_{1}^{2}-mp_{2}q_{1}r_{2}^{2}\,,\\
w&=q_{1}q_{2}\left(nk_{L}+mk^{2}-mk_{L}k_{R}\right)+p_{2}q_{1}\left(2ml_{2}kr_{2}+mk_{L}r_{2}^{2}-l_{2}^{2}n+l_{2}^{2}mk_{R}\right)\\&+p_{1}q_{2}\left(l_{1}^{2}mk_{R}+2l_{1}mkr_{1}+mk_{L}r_{1}^{2}-l_{1}^{2}n\right)-mp_{1}p_{2}\left(l_{1}r_{2}-l_{2}r_{1}\right)^2\,.
\end{split}
\end{equation}
Here, the subscript $i=1,2$ should be regarded modulo 2. 
\section{Fusion of $\T$ defects}
\label{app: T fusion}
In \cite{Fuchs:2007tx} (see also \cite{Thorngren:2021yso}) it was shown that the $\mathcal{T}$ defect is non-invertible for $N\neq 1$ and squares to a sum of $U(1)_m$ defects. Let us now compute its fusion rule (we consider here the case of integer $R^2=N$) using the defect Lagrangian.

Consider two $\mathcal{T}$ defects at positions $S$ and $S+\d S$, and let $S^I$ be the region of spacetime between those defects. The action is then 
\begin{equation}
\begin{split}
    S = & \,\frac{N}{4\pi} \int_{S^-}   d\f_L \wedge \star  d \f_L  + \frac{iN}{2\pi} \int_{S} \f_L d\f_I 
    +\frac{N}{4\pi} \int_{S^I}   d\f_I \wedge \star  d \f_I \\
    &+ \frac{iN}{2\pi} \int_{S+\d S} \f_I d\f_R
    + \frac{N}{4\pi} \int_{S^+}   d\f_R \wedge \star  d \f_R \,,
\end{split}
\end{equation}
and taking the limit $\d S \rightarrow 0$ we arrive at 
\begin{equation}
   S = \frac{N}{4\pi} \int_{S^-}   d\f_L \wedge \star  d \f_L  + \frac{iN}{2\pi} \int_{S} d \f_I (\f_L - \f_R) +\frac{N}{4\pi} \int_{S^+}   d\f_R \wedge \star  d \f_R  ~. \label{T fusion}
\end{equation}
We can rewrite the above as  
\begin{equation}
   S = \frac{N}{4\pi} \int_{S^-}   d\f_L \wedge \star  d \f_L  +  \frac{i}{2\pi}\int_{S} d\varphi (\f_L - \f_R) +\frac{i }{N}\,\eta \int_S d\varphi +\frac{N}{4\pi} \int_{S^+}   d\f_R \wedge \star  d \f_R~,  
\end{equation}
where $d\varphi$ is a compact scalar $\varphi \sim \varphi + 2\pi$ and $\eta = 0,1,\dots, N-1$ is an integer-valued field. To see that, notice that the path integral over $\eta$ is just the sum
\begin{equation}
    \int D\eta \, e^{\frac{ i }{N}\,\eta \int_S d\varphi } = \sum_{\eta=0}^{N-1} e^{\frac{i }{N} \eta \int_S d\varphi } \,,
\end{equation}
which imposes  
\begin{equation}
    \int_S d\varphi \in 2\pi N \mathbbm{Z} \,.
\end{equation}
Redefining $\varphi= N \f_I$ we arrive back to \eqref{T fusion}. All in all, we get that the fusion of two $\mathcal{T}$ defects is described by the action 
\begin{equation}
   S = \frac{N}{4\pi} \int_{S^-}   d\f_L \wedge \star  d \f_L  +  \frac{i }{2\pi}\int_{S} d\varphi \left(\f_L - \f_R  +\frac{2\pi}{N} \eta\right) +\frac{N}{4\pi} \int_{S^+}   d\f_R \wedge \star  d \f_R  ~,  \label{T fusion result}
\end{equation}
with $\eta = 0,1,\dots, N-1$. 

On the other hand, let us consider the insertion of a $U(1)_m$ defect with parameter $\b$. The action is 
\begin{equation}
    S = \frac{N}{4\pi} \int_{S^-\cup S^+}  d\f \wedge \star  d \f + \b \frac{N}{2\p} \int_S \star d\f \,,
\end{equation}
and introducing a Lagrange multiplier $\varphi$ it can be rewritten as 
\begin{equation}
    S = \frac{N}{4\pi} \int_{S^-}  d\f_L \wedge \star  d \f_L + \b\frac{N}{2\p} \int_S  \star d\f_L+ \frac{i}{2\pi}\int_S d\varphi (\f_L -\f_R) +\frac{N}{4\pi} \int_{S^+}  d\f_R \wedge \star  d \f_R \,.
\end{equation}
Since on $S$ we have the relation $d\f_L=d\f_R$ we can write the second term using either fields. We can now remove that term by shifting $\f_L$ by a step function as follows, 
\begin{equation}
    \f_L \rightarrow \f_L -\b\Theta(S^+) \,.
\end{equation}
However, while the change of the first term cancels with the second term, the third term becomes 
\begin{equation}
    \frac{i}{2\pi}\int_S d\varphi \left(\f_L -\f_R- \b \right) \,.
    \label{U(1)t}
\end{equation}

In conclusion, comparing \eqref{T fusion result} with \eqref{U(1)t} we see that the product of two $\mathcal{T}$ defects is a sum of $U(1)_m$ defects, 
\begin{equation}
    \mathcal{T} \times \mathcal{T} = \sum_{\eta=0}^{N-1} e^{ -\frac{2\pi \eta}{N} \frac{N}{2\pi} \oint \star d\f}~.
\end{equation}
Note that for $N=1$ it becomes invertible.

Let us comment that this calculation also explains a statement we made below equation \eqref{trivial}, according to which multiplying \eqref{trivial} by an integer $\kappa$ does not result in a new simple defect but rather in a sum of $U(1)_m$ defects. Indeed, as shown above, the defect Lagrangian 
\begin{equation}
\frac{i\kappa}{2\pi} d\varphi (\phi_L - \phi_R)
\end{equation}
can be written in the form 
\begin{equation}
\frac{i }{2\pi}\int_{S} d\varphi \left(\f_L - \f_R  +\frac{2\pi}{\kappa} \eta\right) \,,
\label{defect Lagrangian eta}
\end{equation}
where $\eta = 0,1,\dots, \kappa-1$ is an integer-valued field. Then, the defect Lagrangian \eqref{defect Lagrangian eta} is just a sum of $U(1)_m$ defects, 
\begin{equation}
\sum_{\eta=0}^{\kappa-1} e^{ -\frac{2\pi \eta}{\kappa} \frac{R^2}{2\pi} \oint \star d\f}~.
\end{equation}
\bibliographystyle{jhep}
\bibliography{refs}

\end{document}